# Forcing Seasonality of influenza-like epidemics with daily Solar resonance

Fabrizio Nicastro[1*], Giorgia Sironi[2], Elio Antonello[2], Andrea Bianco[2], Mara Biasin[3], John R. Brucato[4], Ilaria Ermolli[1], Giovanni Pareschi[2], Marta Salvati[5], Paolo Tozzi[4], Daria Trabattoni[3], Mario Clerici[6]

[1] *Italian National Institute for Astrophysics (INAF) – Rome Astronomical Observatory, Rome, Italy*
[2] *Italian National Institute for Astrophysics (INAF) – Brera Astronomical Observatory, Milano/Merate, Italy*
[3] *Department of Biomedical and Clinical Sciences L. Sacco, University of Milano, Milano, Italy*
[4] *Italian National Institute for Astrophysics (INAF) – Arcetri Astrophysical Observatory, Firenze, Italy*
[5] *Regional Agency for Environmental Protection of Lombardia (ARPA Lombardia), Milano, Italy*
[6] *Department of Pathophysiology and Transplantation, University of Milano and Don C. Gnocchi Foundation, IRCCS, Milano, Italy*

**Summary**
**Seasonality of acute viral respiratory diseases is a well-known and yet not fully understood phenomenon. Several models have been proposed to explain the regularity of yearly recurring outbreaks and the phase-differences observed at different latitudes on Earth. Such models take into account known internal causes, primarily the periodic emergence of new virus variants that evade the host immune response. Yet, this alone, is generally unable to explain the regularity of recurrences and the observed phase-differences.**
**Here we show that seasonality of viral respiratory diseases, as well as its distribution with latitude on Earth, can be fully explained by the virucidal properties of UV-B and A Solar photons through a daily, minute-scale, resonant forcing mechanism. Such an induced periodicity can last, virtually unperturbed, from tens to hundreds of cycles, and even in presence of internal dynamics (host's loss of immunity) much slower than seasonal will, on a long period, generate seasonal oscillations.**

## Introduction

Several models have been proposed to explain the regularity of yearly recurring outbreaks of acute respiratory diseases, and their phase-differences at different latitudes on Earth (e.g. Lofgren et al., 2007 and references therein). Internal biological dynamics that allow influenza viruses to evade host's immunity and become more virulent (Cox and Subbarao, 1999), manifest mainly through "antigenic drift", the property of such viruses to mutate rapidly (Agranovski et al., 2004; 2005). For A and B influenza viruses, such mutations develop at rates of $6.7 \times 10^{-3}$ and $3.2 \times 10^{-3}$ nucleotide substitution per site per year, respectively (Logfren et al., 2007; McCullers et al., 1999). These imply, on average, 6 and 3 mutations per RNA segment of each virus every 6 months, respectively, sufficient (especially for type-A influenza) to trigger recurrences of epidemics on antigenic-drift timescale. However, without an external forcing mechanism, such recurring events are typically of modest intensity, can only last for a few cycles, and most importantly are not able to reproduce the observed phase-shift between north and south hemisphere on Earth.

One or more concurring external mechanisms must be invoked to explain all evidence, and several have been proposed (e.g. Ebi et al., 2001; Grais et al., 2003; Grais et al., 2004; Liao et al., 2005; Viboud et al., 2004; Baker et al., 2020). These include, but are not limited to, weather-related temperature and humidity oscillations on Earth (e.g. Baker et al., 2020), the observed yearly pattern of air-travel (e.g. Grais et al., 2003, but see also Grais et al., 2004), indoor heating during winter (e.g. Liao et al., 2005), bulk aerosol transport of virus through global convective currents or related to weather oscillations like "El Niño" (e.g. Ebi et al., 2001; Viboud et al., 2004).

Most of these, however, modulate on timescales that are either typical of seasons on Earth or longer and thus singularly can only explain some aspects of seasonality of recurring outbreaks, or the onsets of particularly violent outbreaks. For example, the load of air-travel flow is typically larger during northern hemisphere summers, and this is in contrast with most of the outbreaks developing during falls and winters in these countries. On the other hand, vastly populated areas in the north hemisphere lie at relatively low latitudes in temperate regions where indoor heating is not massively widespread: yet, even in these areas, influenza epidemics typically develop during winters, and remain inactive in summers. Similarly, temperature, humidity, as well as air pollution, are not homogeneously distributed through the planet, and micro-climate can vary substantially at the same latitude, depending on the particular geographical or industrial setting of even contiguous areas. The cross-influence of all these factors could certainly synergize to produce the observed geographical and time modulations, but this would require a hard to obtain perfect fine-tuning of the many parameters at play.

The one mechanism that has been acting constantly at (almost) the same pace every day of the year and every year for the past four billion years, is the irradiation of the Sun on Earth. At a given location on Earth, this is modulated daily by the Earth's spinning and yearly by the Earth's orbit around the sun (Figure 1). Solar ultraviolet photons with wavelength in the range 200-290 nm (UV-C radiation) photo-chemically interact with DNA and RNA and are endowed with germicidal properties that are also effective on viruses (Kowalski, 2010; Rauth, 1965; Kesavan and Sagripanti, 2012; Chang et al., 1985; McDevitt et al., 2012; Welch et al., 2018; Walker and Ko, 2007; Tseng and Li, 2005). Fortunately, UV-C photons are filtered out by the Ozone layer of the upper Atmosphere, at around 35 km (e.g. Walker and Ko, 2007). Softer UV photons from the Sun with wavelengths in the range 290-320 nm (UV-B) and 320-400 nm (UV-A), however, do reach the Earth's surface. The effect of these photons on Single- and Double-Stranded RNA and DNA viruses (e.g. Lytle and Sagripanti, 2005; Lubin and Jense, 1995) and the possible role they play on the seasonality of epidemics e.g Martinez, 2018), are nevertheless little studied. In two companion papers, we present a number of concurring circumstantial pieces of evidence suggesting that the evolution and strength of the recent Severe Acute Respiratory Syndrome (SARS-Cov-2) pandemics (Zhu et al., 2019; Cobey, 2020), might have been modulated by the intensity of UV-B and UV-A Solar radiation hitting different regions of Earth during the diffusion of the outbreak between January and July 2020 (Nicastro et al., 2020, submitted), and provide a measurement of the lethal UV-C dose for Covid-19 (Bianco et al., 2020, submitted).

For the solar virus-inactivation mechanism to be effective in modulating the disease's transmission and settling the duration and recurrence of epidemics, both direct contact (e.g. through exposed surfaces) and airborne transmission have to play a role. This is indeed the case for most respiratory diseases (e.g. Prather et al., 2020 and references therein). Additionally, the time needed for aerosol particles to gravitationally settle has to be comparable or longer than the typical UV-B/A virus's lethal time (Nicastro et al., 2020, submitted). This is probably not true (in still air) for the largest (~100 μm in size) and thus heaviest droplets produced in coughs and sneezes, where the gravitational settling time is of the order of few seconds (Morawska and Cao, 2020), but can certainly be the case for significantly smaller, micron or sub-micron size, aerosol particles (like those that commonly contain the influenza virus, e.g. Tellier et al., 2019) that can survive in still air for longer than 12.4 hours (Tellier et al., 2019).

Here we present a simple mathematical model of epidemics that includes the effect of Solar UV photons ("solar-pump", hereinafter) as an infra-day resonance mechanism that acts systematically every day with strength and on timescales (from minutes to hours) dictated only by astronomical parameters, and explore its parameter space. We demonstrate that such a physically-motivated model is able to naturally reproduce the seasonality of influenza-like epidemics as well its distribution with latitude on Earth, without forcing seasonal oscillations with ad-hoc mathematical prescriptions (e.g. Dushoff et al., 2004; Neher et al., 2020). Moreover, it matches reasonably well the Italian data of the recent SARS-CoV-2 epidemics (see Methods in Supplemental Information: SI), predicting different short-term future scenarios, depending on the efficiency of the solar-pump.

## Results
**The Model**
Our model considers isolated (i.e. no cross-mixing), zero-growth (i.e. death rate equal to birth rate), groups of individuals in whom an efficacious anti-viral immune responses has either been elicited (Susceptible → Infected → Recovered: SIR) or not (Susceptible → Infected → Recovered → Susceptible: SIRS). The inefficiency of the anti-viral immune response is parameterized through a population loss-of-immunity (LOI) rate $\gamma_{in}$ (zero for no LOI). Other parameters of the model are the rate of contacts β (proportional to the reproductive parameter $R_t$, with $R_{t=0} = R_0$ being the intrinsic reproductive number of the epidemic), the rate of recovery $\gamma_{out}$, the rate of deaths μ, lockdown/phase-2 halving/doubling times, when considered (see below), and the efficiency ε of the solar-pump (see below).

*The solar-pump*

In our equations (see Methods in SI) the solar-pump forcing term is introduced as an infra-day (minutes/hours) attenuation factor for the reproductive number $R_t$ and modeled by following the sun radiation curves as a function of Earth's latitude, day of the year, and time of the day. Solar curves (Fig. 1, and onsets of Fig. 2) are functions of the solar height angle α (complement to the Zenith angle: i.e. α=90 degrees at local noon) that, in turn, depends on Earth's latitude φ, sun's declination δ, day of the year $d$ and time of the day $h$. Solar curves modulate daily with $\sin(\alpha)$ and have maximum amplitudes and widths that depend on φ, $d$ and $h$, as shown in Fig. 1, where the $R_t$ attenuation factor $(1 - \varepsilon \sin(\alpha))$ is plotted as a function of $h$ (see figure's caption for details).

The efficiency of the solar-pump is here parameterized through the parameter ε (in the interval [0,1]) but depends, effectively, on internal virus properties, namely the lethal time $\tau_0$ needed to deliver the UV lethal dose $D_0$ (defined as the UV-C fluence needed to inactivate 63% of the virus) for the given virus (Nicastro et al., 2020, submitted). At local noon (the minima of the daily solar curves of Fig. 1) typical UV-B/A solar intensities on Earth range from 0.9–18 J m$^{-2}$ min$^{-1}$ in the north and south hemispheres between winters and summers, respectively, and are roughly constant at the equator throughout the year with a mean value of ~ 18 J m$^{-2}$ min$^{-1}$. Folding these intensities through the action spectrum (i.e. the ratio of the virucidal dose of UV radiation at a given wavelength to the lethal dose at 254 nm) of Lytle and Sagripanti (2005), yields to lethal UV-B/A intensities of ~0.05–1 J m$^{-2}$ min$^{-1}$. For Covid-19 we measured a three orders of magnitude inactivation dose of 37 J m$^{-2}$ (Bianco et al., 2020, submitted), which translates into $D_0$=4.8 J m$^{-2}$. This value is used to produce the lethal-time dotted curves of Figures 3–9, which oscillate between 5-100 minutes at local noon depending on latitude and time of the year, and corresponds to our reference solar-pump efficiency ε=1. Efficiencies lower than the reference value 1 correspond to inversely longer lethal-times (e.g. dotted green curve of Fig. 9). As an example, Figure 2 shows two solar-pump attenuated $R_t$ curves over a solar-year period for a 40 degree north location on Earth and for an epidemic with $R_0$=3 (see figure's caption for details).

*Lockdown and Phase-2*

Lockdown and phase-2 terms are modeled with exponential functions (see Methods in SI) that act as attenuation (lockdown) and amplification (phase-2) factors of the reproductive number $R_t$, with, respectively, halving and doubling times (in units of 1 day) that are free parameters of the model. The lockdown attenuation ends when phase-2 starts, and phase-2 amplification ends when $R_t$ reaches the original value $R_0$ characteristic of the disease.

**Our two Study-Cases**

We focus on two particular study-cases: (A) typical influenza epidemics, with initial reproductive number $R_0$=1.5, solar-pump efficiencies ε =1 and 0 (i.e. no solar-pump) and recovery and death rate $\gamma_{out}$=0.2 and μ=0.001, respectively, and (B) a more aggressive and higher-mortality epidemics, with parameters similar to those recorded for the recent SARS-Cov-2 pandemics ($R_0$=3, $\gamma_{out}$=0.1 and μ=0.01) and solar efficiencies ε=1, 0.3 and 0. For each case, we consider three different LOI rates: $\gamma_{in}$=0 (i.e. no LOI), $\gamma_{in}$=0.0055 (180-day period: shorter than seasonality) and $\gamma_{in}$=0.00055 (5-year period: much longer than seasonality).
Finally, for each case, we consider three groups of individuals with equal populations (sixty million) and living at three different Earth's latitude: 40 degrees north (black curves in all figures), 40 degree south (blue curves in all figures) and the equator (orange curves in all figures). Groups are taken as isolated and have no cross-relation between them.

*Case A: Influenza Epidemics*

Figure 3 shows the curves of growth of the infected individuals in the three latitude groups over the first 20 years of influenza epidemics with (Fig. 3a) and without (Fig. 3c) solar-pump (see caption for details). When the solar-pump is not active (ε=0; Fig. 3c) the growth (and decline) of epidemics proceeds identically at all latitudes (black, blue and orange curves are superimposed to each others). The period of oscillations, in this case, resonates solely with the inverse of the LOI rate (the time it takes for previously infected individuals to become again susceptible), the strength of epidemics declines from cycle to cycle, and the disease dies out completely in less than 10 years. When the solar-pump is active (ε=1; Fig. 3a), instead, the reproductive number undergoes small infra-day variations (whose exact magnitude depends on the particular time of the year and Earth's latitude, other than the efficiency of the solar-pump mechanism: e.g. onsets of Fig. 2) that do not significantly change its daily average value (Fig. 2), but progressively drive the modulation of epidemic oscillations to match the seasonal solar period (Fig. 3a). The north (black), south (blue) and tropical (orange) groups differentiate immediately, and after a few cycles (lasting a total of about 5 years) are perfectly regulated on biannual periodicity, with phase shifted in time by 6 months in the north versus south hemisphere and the tropical group undergoing fast and continuous moderate bursts of epidemics roughly every six months. The biannual oscillations at latitudes ±40 degrees, last for a few tens of years, and then synchronize with the yearly solar-irradiation seasonal period (see figure's caption for details). The cycles can run unperturbed for centuries, as shown in the Susceptible versus Infected (S–I) phase-diagrams of Fig. 3b, where the first 100 years of endemic diseases are drawn (see Figure's caption for details).

Fig. 4 and 5 show analogous figures for the cases with $\gamma_{in}$=0.0055 (5-year period of recurrence: Fig. 4) and $\gamma_{in}$=0 (no LOI: Fig. 5). Interestingly, when the solar-pump is active and LOI periods are long compared to the seasonal period (Fig. 4a and 4b), oscillations are more erratic and chaotic during the long settling period (initial cycles cover a wide dynamical range in the S–I phase diagram of Fig. 4b). Destructive resonance between the LOI rate and the Solar cycles acts differently in the three latitude groups, and if no group mixing is introduced (and with this particular set of model parameters), the intense bursts of epidemics in the north hemisphere are initially separated by the inverse of LOI rate (Fig. 4a). This separation progressively squeezes at all latitudes as the continuous solar-pumping imposes its rhythm, and becomes strictly seasonal at all latitudes in about 40-50 years. The squeezing of the time intervals between outbreaks, is visible also when the solar-pump is not active (Fig. 4c). This is because, as new outbreaks develop, the reservoir of susceptible individuals made available by previous LOI cycles and the number of people still infected at any time are sufficient to trigger new bursts at earlier times. However, in this case, the strength of each outbreak decreases smoothly and systematically with years, and in about 70 years the reservoir of susceptible individuals at each time completely empties and the epidemics die out (Fig, 4c and 4d). Moreover, again, the three latitude groups behave identically, which is not observed.

Finally, when $\gamma_{in}$=0 (no antigenic-drift), the solar-pump cannot, alone, sustain stationary epidemic cycles, but strongly modulates the course of epidemics during the first three years in the north and south hemisphere and over a longer period of about 8 years in the tropical area (Fig. 5a). By contrast, when the solar-pump is not active (Fig. 5c) the epidemics develop identically at all latitudes and last for a single 1-year cycle (Fig. 5d).

*Case B: SARS-CoV-2 Pandemics*

Our second study-case aims to model the recent SARS-CoV-2 pandemics outbreak under the different assumptions tested in this work, and allows us to gain insights on possible future scenarios. As for study-case A, we use the three groups at latitudes $40^0$-north, $40^0$-south and the equator, and run simulations both with ($\varepsilon\neq0$) and without ($\varepsilon$=0) the solar forcing term and for LOI rates $\gamma_{in}$=0.0055, 0.00055 and 0. Additionally, for study-case B we also compare our north-hemisphere simulation with the SARS-CoV-2 data for Italy (central latitude +41.5 degrees) observed from 24 February 2020 through 21 August 2020 (see Methods in SI). To do so, we introduce, for all groups, two additional external forcing terms that modulate the contact rate β (chosen to be 0.33, to reproduce the observed – in virtually every monitored country – initial exponential growth rate): a 'lockdown' term and a 'phase-2' term (see Fig. 6's caption for details).

With the particular initial setting described in the caption of Fig. 6, the simulation with active solar-pump ($\varepsilon$=1) and fast antigenic-drift ($\gamma_{in}$=0.0055) reproduces extremely well the curve of growth of the observed daily number of infected in Italy (Fig. 6a, black infected curve and data), with an under-sampling factor of 6 (i.e. observed infected are one sixth of actual infected). Preliminary results from serological tests on a statistically selected sample, confirm this average under-sampling factor in Italy (see e.g. https://www.slideshare.net/slideistat/primi-risultati-dellindagine-di-sieroprevalenza-sarscov2).

Fig. 6c and 6d shows the same simulation when the solar-pump is not active ($\varepsilon$=0; curves of growth from the three latitude groups are superimposed to each other). However, with these parameters, the under-sampling factor needed to fit the initial exponential phase of the growth of contagions in Italy is an unrealistic factor of 300. Moreover, the declining phase of contagions, with no solar-pump is too fast to reproduce the Italian data. The much smaller absolute number of daily infected in the simulation shown Fig. 6a, compared with that of Fig. 6c (under-sampling factor of 6, versus 300), and the slower decay of daily detected infections, are due solely to the daily effect of the solar-pump. The fast-decay in the inactive solar-pump case is a direct consequence of the much larger 'consumption' of susceptible individuals during the growth of contagions in the first cycle (compare Fig. 6d with Fig. 6b).

The sun has also the effect of reducing by almost one and two orders of magnitude, respectively, the number of infected individuals observed from February through May, in the South and tropical groups, with respect to the North group, which is exactly what has generally been observed during the SARS-CoV-2 pandemic (Nicastro et al., 2020, submitted; Fig. 6a). The curve of growth of the South group, however, starts climbing again in March–April (which has generally been observed, e.g. Nicastro et al., 2020, submitted) and, if no additional distancing measures are adopted, should have a second and higher peak in October, towards the end of the South winter (Fig. 9a). In the north hemisphere, instead, this simulation (but see below for alternative scenarios) predicts that the curves of growth of the epidemics reach a minimum in June–July (as indeed observed in Italy: Fig. 6a), start climbing again (less aggressively than during the initial January outbreak) from August through October and, in absence of new distancing measures, reach a new intense peak in November, while tropical countries will see a second and much stronger outbreak during the second half of 2020 and through January–February 2021. Subsequently, under these conditions, the epidemics will start adjusting to the solar cycle and less aggressive outbreaks (maximum daily number of new infected individuals between ten and forty thousand) will repeat cyclically for tens of years (Fig. 6b).

Should recovered people preserve their immunity against Covid-19 ($\gamma_{in}$=0 case), UV photons from the Sun may still be able to indirectly amplify contagions during autumns/winters (as a consequence of the attenuation during previous summers) and produce one additional wave of the SARS-CoV-2 epidemics at all latitudes (compare Fig. 7a and 7b with Fig. 7c and 7d, respectively). However, in about two years (and only 1 additional cycle), in absence of antigenic-drift or cross-mixing between groups at different latitudes, the epidemics will die out (Fig. 7b).

Finally, as for study-case A (influenza), at the much slower LOI rate $\gamma_{in}$=0.00055 (LOI period of 5 years), the seasonal cycle takes much longer to stabilize, and oscillations proceed at intermittent intensities both in the north and south hemisphere (with different phases) for tens of years, and then finally synchronize to the Solar cycles on timescales longer than 50 years (Fig. 8).

## Discussion

Our model is able to reproduce in a particularly precise way the seasonality of respiratory diseases where airborne transmission is the main path of infection and their phase-latitude dependence. The solar-pump mechanism works more efficiently on free-to-spread (i.e. no distancing measure adopted) epidemics with intrinsic reproductive number $R_0$ lower than ≈2–3. For epidemics with larger intrinsic reproductive number, and no restriction measures at work, the first outbreak could quickly produce heard immunity (e.g. Baker et al., 2020) and the onset of seasonality due to the solar-pump mechanism, may be harder to achieve even in presence of fast LOI rates. Nonetheless, even for highly contagious diseases that turn into pandemic, like the recent SARS-CoV-2, our model is able to explain the qualitative differences generally observed when the strength of the pandemics is compared in countries located in the north, tropical and south hemisphere (e.g. Nicastro et al., 2020, submitted).

Accurate predictions of ongoing epidemics depend critically on the initial and boundary conditions of the simulations. For example, the particular set of model parameters we used to match the Italian data of the ongoing SARS-CoV-2 epidemics (see Fig. 6's caption for details), is obviously not unique. In particular, these predictions depend (given the initial dates for lockdown and phase-2) on the starting time of the epidemics and the actual efficiency of the solar effect (two parameters strongly degenerate with each others: see below). The second of these parameters, in turn, depends on the exact knowledge of the virus' UV-C lethal dose $D_0$, the way this is delivered and the time the virus (in aerosol or surfaces) remains exposed to the dose. Accurate modeling of this phenomenon is beyond the scope of this work. However, it is useful to show the comparison with drastically different solar-pump efficiencies to highlight the potential predictive power of the model for e.g. the SARS-CoV-2 epidemics.

Fig. 9 shows the results of three simulations for the group at 40 degrees north, and solar-pump efficiencies ε=1 (same as Fig. 6a; red curve), ε=0.3 (~3 times lower than in the case of Figure 6: compare, e.g. green and blue curves of Fig. 2; green curve) and ε=0 (blue curve). Some of the remaining model parameters are tuned to match the Italian SARS-Cov-2 curve of growth of contagions (see figure's caption for details) from the start of epidemics through 21 August 2020 and with the observed under-sampling factor of 6. Most of the tuned parameters can be compared with observables, which allows for a quantification of the solar-pump effect. In particular, the red, green and blue curves have epidemic starting dates $t_0$=6, 19 and 22 January 2020. This parameter is particularly critical in setting the normalization of the curves and the date of their first peak, given the known under-sampling factor of 6 (as observed in Italy) and lockdown starting date of 11 March 2020 (the official date of lockdown in Italy). For comparison, the first Covid-19 cases have occurred in Italy over the 1–10 January 2020 period (Cereda et al., 2020). Two additional sensible parameters, which model the shape of the peak, the initial declining phase of the first outbreak and the recovery phase of contagions, are the $R_t$ halving time during lockdown and the phase-2 starting date. The plotted models have halving times of 28 (red curve), 18 (green curve) and 15 days (blue curve). For comparison, in the three regions of Italy most heavily hit by the outbreak and that largely dominated the number of daily infected till July 2020, $R_t$ halving times from their peak values have been estimated in 24 (Lombardy), 17 (Emilia-Romagna) and 28 days (Veneto; see Flavia et al. 2020 and Moirano et al., 2020). Finally, phase-2 starting dates in the models are 17 (red curve), 10 (green curve) and 12 (blue curve) April 2020, respectively 9, 16 and 14 days earlier than the official start date of phase-2 in Italy. These comparisons suggest a relatively high solar-pump efficiency during the first six months of the SARS-CoV-2 epidemic in Italy.

An additional way of assessing the relative importance, if not the exact magnitude, of the solar-pump mechanism versus other external forcing mechanisms, will be to compare the evolution of the epidemic in Italy over the next several months with model predictions. The three curves of Fig. 9 differ substantially in short and long-term predictions of the evolution of the epidemic. In all cases, Italy, where distancing measures have now greatly being relaxed, should start seeing a strong increase in the number of new daily infected in September, and this should continue, in absence of distancing measures, through the end of November 2020 for maximum solar-pump efficiency (red curve) and through mid January 2021 for low solar-pump efficiency (ε≤0.3: green and blue curves). Moreover, at low solar-pump efficiencies (green and blue curves) the number of daily infected shall start climbing again around late spring 2021 and then again, roughly every six months for a few additional cycles, dictated only by the LOI rate ($\gamma_{in}$=0.0055 in these simulations) and until the solar mechanism starts differentiating the green from the blue curve. On the contrary, at maximum solar-pump efficiency (red curve) contagions will not be able to spread again till late summer 2021, and then will adjust to seasonal cycles.

Finally, as shown in Fig. 6–8, at each given given time the solar-pump mechanism acts differently at different latitudes on Earth. Thus, comparing our model predictions with SARS-CoV-2 data in Countries located at opposite latitudes over the next few months, should also tell us quantitatively about the importance of the solar-pump effect in modulating the spreading of this disease.

## Limitations of the Study

Our model for the diffusion of contagions (see Methods in SI) considers population groups as ideal isolated systems (i.e. no cross-mixing of populations in different groups at different latitudes on Earth). This condition is certainly not verified in today's global world, in which daily exchange of people and goods are massive and, generally, not strictly controlled (unless measures to limit the spread of contagions are taken by governments, as in the case of the recent SARS-CoV-2 pandemics). This type of complexity could easily be introduced in our basic system of differential equations (see Methods in SI) by considering cross-mixing through a matrix of either observed (i.e. based on air- ground- and sea-traffic data) or randomized exchange rates of susceptible and infected individuals among groups. However, the introduction of such an additional degree of complexity in the model (which will be the subject of a forthcoming paper), while could contribute to wash-out and partially homogenize differences between specific groups at different latitudes, could hardly build-up or break the strict seasonality that is typically observed in the recurrence of respiratory diseases and whose explanation is the main scope of this work.

The basic ingredient of our model that generates seasonality and latitude-dependent phase differences among groups, is a physically-motivated resonant forcing mechanism (the solar-pump) that acts as a daily modulator of the reproductive number $R_t$ (see Methods in SI). The mechanism is based on the virus-inactivation power of the solar UV photons that reach the Earth's surface, and is dictated by the shapes of the daily solar irradiation curves on Earth, which, for a given latitude and time of the year, depend only on the Earth's spinning around its rotation axis and whose self-similarity throughout a solar year is determined only by Earth's orbit around the Sun. However, the exact efficiency of this mechanism depends on the quantification of several intrinsic parameters of the specific epidemic, such as the lethal dose of UV photons needed to inactivate the specific virus when exposed to solar-like spectra, the average life-time of the virus in aerosol or surfaces, its transport in wind or low-atmosphere currents, etc. The quantification of these parameters for given viruses, is beyond the scope of this work, whose aim is to show the potential of solar radiation in modulating the short- and long-term evolution of epidemics outbreaks and demonstrate that the, unavoidable, infra-day (minute timescales) resonant mechanism naturally offered by the solar-pump, can efficiently synchronize with internal LOI rates to explain the course of epidemics on much longer timescales.

Related to the above caveat, the efficiency of the solar-pump mechanism, which can only act on outdoor droplets generated through speaking, coughing and/or sneezing, aerosol and surfaces-deposited particles, depends also on the social habits of different population groups, located even at similar latitudes on Earths, often dictated by climate, and on specific location's weather (e.g. diffuse cloud coverage). For example, outdoor living, even in winters, is much more common in Mediterranean countries than, e.g. in North-American states at similar Earth's latitudes. This is often anti-correlated with the massive use of indoor air-conditioning (offices, shopping malls, private homes, private and public transportation, etc.), which is known to be an efficient mean of diffusion of contagions and could, e.g., contribute to explaining why the SARS-CoV-2 epidemics has spread aggressively during the initial part of summer in northern locations like the state of Texas in the US. Similarly, the diffuse cloud coverage (especially during the times of the days when the solar irradiation is maximum) in tropical countries, during specific times of the year (e.g. the monsoon season in India, or the rain seasons in the northern of Brazil, or hurricane seasons in central and tropical America) will reduce the efficiency of the solar-pump and could act as an amplifier of the growth of contagions in specific periods of the year. Again, the consideration of such additional degrees of complexity, in the model is certainly possible but would require a somewhat arbitrary (if not based on solid social and weather data) introduction of group-dependent weights in the parameterization of the efficiency of the solar-pump mechanism, which is beyond the scope of this paper.

Finally, we stress that the large number of parameters of the model together with the degeneracy of some of these parameters, make it difficult to exactly quantify the relative importance of the solar-pump mechanism in epidemics with high intrinsic reproductive number and for which eternal forcing mechanisms like lockdown and phase-2 are introduced. As discussed in the session above, the epidemics starting time in the model is strongly degenerate with the under-sampling factor (the ratio between actual infections and detected cases), the efficiency of the solar-pump and, more moderately, the starting date of adoption of distancing measures. The exact shape of the declining and recovery phases of the curves of growth of the epidemics, when distancing measures are in effect and after they are relieved, also influence the proper modeling of actual epidemics data. In our model we assume that during this two phases, the contact rate (and so $R_t$) evolves exponentially, with one single halving and doubling time, respectively (see Methods in SI). This does not have to be (and most likely is not) the case, practically. Measures adopted can (and most often do) vary on the go, and this implies either different functional shapes or variable halving/doubling times within the same phase. Such complexity introduces degeneracies between the exponential halving and doubling times and the starting dates of lockdown and phase-2 in the model, when modeling global data over externally controlled declining and recovering phases. A data-driven modeling approach to the problem, together with an accurate knowledge of all observables of the epidemics, are the only way to break these degeneracies without increasing the degree of complexity of the model (which, in turn, would introduce additional degrees of degeneracy).

## Acknowledgments

The authors thank an anonymous referee who helped to improve the clarity and quality of the paper substantially. The work presented in this paper has been carried out in the context of the activities promoted by the Italian Government and in particular, by the Ministries of Health and of University and Research, against the COVID19 pandemic. Authors are grateful to INAF's President, Prof. N. D'Amico, for the support and for a critical reading of the manuscript. Authors are also grateful to Dr. M. Elvis for critically reading the paper and providing useful comments and suggestions. A research grant from Falk Renewables partially supported this work.

## Author Contributions

F.N. designed the study (together with co-authors P.T., G.P. and M.C.), wrote the program that solves the model's equation and provides the curves of growth of the epidemics, collected and analysed Italian SARS-CoV-2 data and wrote the paper. G.S. and M.S. collected, reduced and analysed TEMIS data. G.P., M.C., M.T., M.B. and A.B., provided the measurement of the UV lethal dose for Covid-19 and, together with co-authors G.S., J.R.B., E.A. and I.E., derived the action-spectrum-folded UV-B/A virus lethal times plotted in Fig. 3–9. All authors contributed equally to the discussion of the results and commented on the manuscript.

## Declaration of Interests

Authors have no competing interest to disclose.

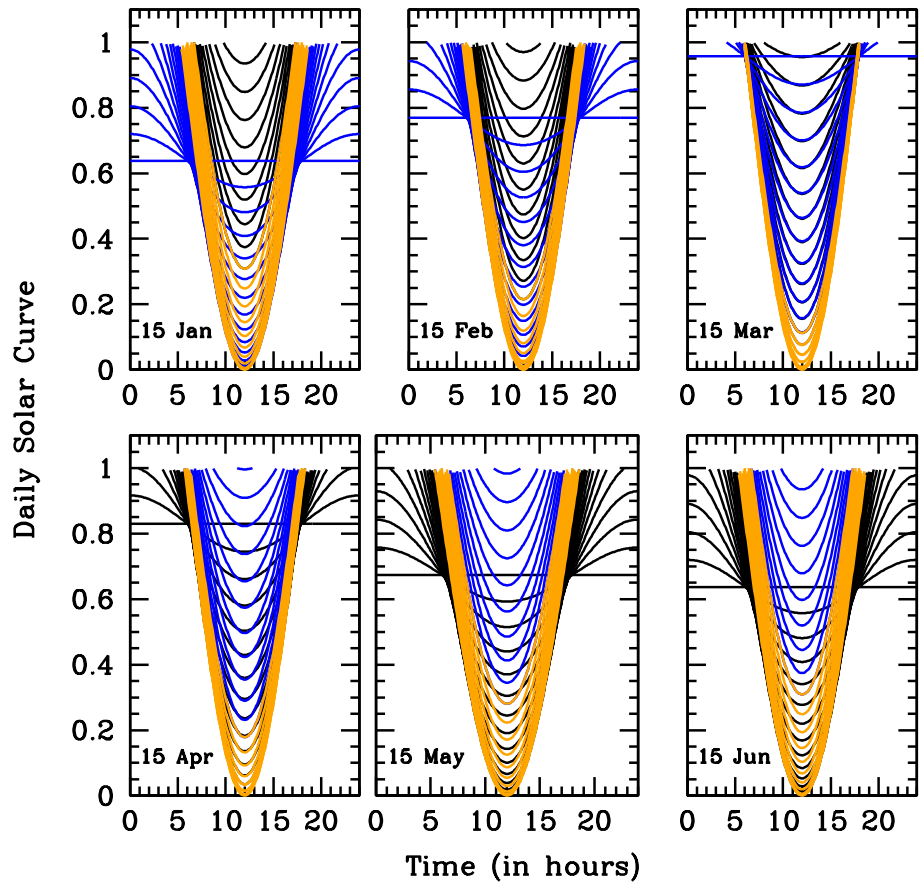

**Figure 1 The Solar Pump.** Attenuation functions for the reproductive number $R_t$, due to the daily solar action. Each panel show solar attenuation curves for 36 Earth's latitudes between -90 and -25 degrees south (blue), -25 – +25 degrees (orange) and 25–90 degrees north (black) at the 15[th] day of each of the first 6 months of the year, as labeled. The plotted function is $(1 - \varepsilon \sin(\alpha))$, where $\alpha$ is the solar height angle (complement to the Zenith angle: i.e. $\alpha$=90 degrees at local noon) that depends on Earth's latitude $\varphi$, sun's declination $\delta$, day of the year $d$ and time of the day $h$. Attenuation functions are defined only for $\alpha>0$ (i.e. during daytime) and are shown for the reference solar-pump efficiency $\varepsilon$=1.

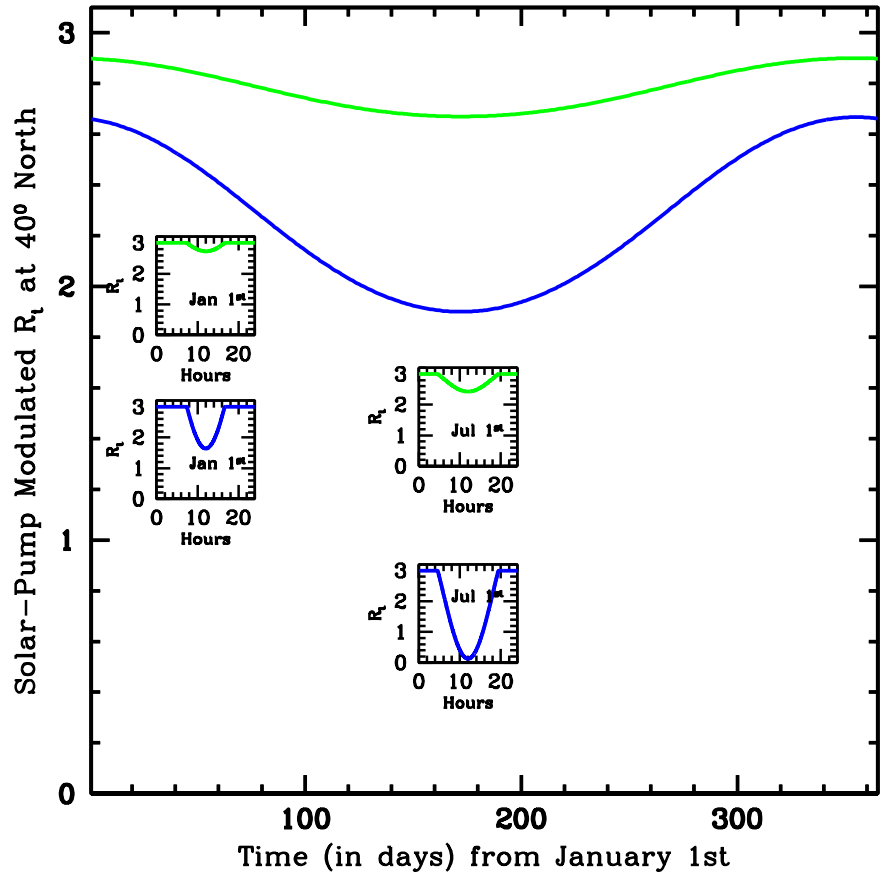

**Figure 2 Efficiency of the solar-pump.** One-day averaged reproductive number $R_t$ as a function of time (in days) over one full solar year and at a 40 degree north location on Earth, attenuated only by the solar-pump mechanism with efficiencies of $\varepsilon=0.3$ (green curves) and $\varepsilon=1$ (blue curves), starting from an epidemic-intrinsic $R_0=3$ value. Onsets show the daily attenuation curves as a function of day-time in hours during the first and the 180[th] day of the year (January 1[st] and July 1[st], respectively).

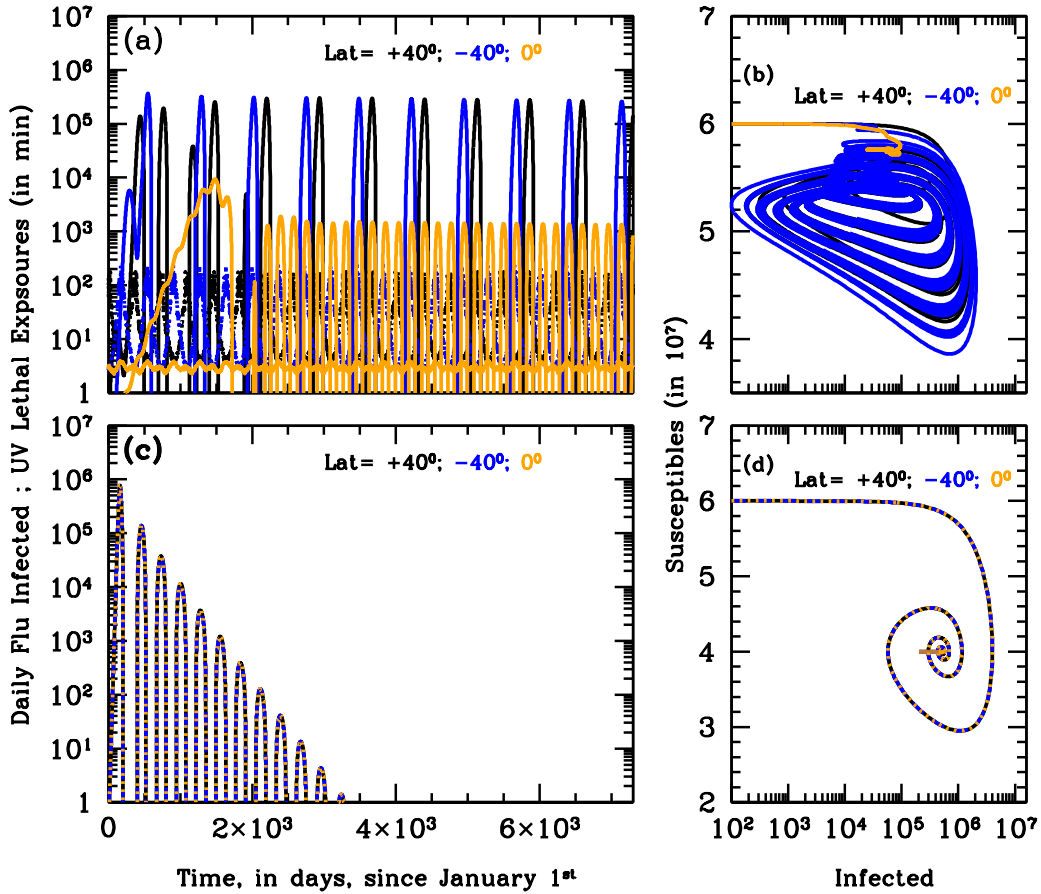

**Figure 3 Seasonality of Influenza.** Panels (a) and (c): curves of growths of the daily new infected for the first 20 years of annual Influenza outbreaks at three different latitudes on Earth, +40 degrees north (black), -40 degrees south (blue) and the Equator (orange), with ($\varepsilon=1$; panel (a)) and without ($\varepsilon=0$; panel (c)) active solar-pump. Model parameters are $t_0$=January 1st, $t_{end}$=t0+100 years, $R_0$=1.5, $\gamma_{out}$=0.2, $\gamma_{in}$=0.0055, $\mu$=0.001 and no external lockdown or phase-2. When the solar-pump is active (panel (a)) constructive resonance builds-in regular sun-modulated oscillations already after the first 2 cycles, and the strength of the seasonal outbreaks adjust to constant values that depend on latitude. With these particular model parameters, north- and south-hemisphere oscillations have initially a 2-year period and are shifted in phase by 6 months (with outbreaks only during local winters), while at the Equator the disease is continuously spread at a low-intensity level throughout the year. It takes almost 50 years before oscillations at ±40 degrees latitudes become annual. This happens at the expenses of the maximum intensity of the biannual outbreaks, which starts declining smoothly as mild outbreaks develop during the initially off-state year. Without solar-pump (panel (c)) the cycles follow the LOI period and are the same at all latitudes: the epidemics quickly die out in a few years. Yearly oscillating dotted curves at the bottom of the top panel are UV-B/A lethal-exposure estimated for Covid-19 and are reported here only to mark winter peaks and summer minima.

Panels (b) and (d): S–I phase diagrams for the simulations of panels (a) and (c), respectively. The first 100 years of the epidemics are plotted (every other 10 years, in panel (b), for easier visualization). When the solar-pump is not active (panel (d)), the curves are the same at all latitudes and the cycles of epidemics die out quickly (essentially in 3-4 cycles) because of lack of supply among the susceptible population. When the solar-pump is switched on (panel (b)) the resonant mechanism builds up seasonality quickly. During the first 50 years, and at latitudes of ±40 degrees (black and blue curves), the susceptible population at each time oscillates between 65–98% of the population, and the instant fraction of infected people over susceptible, at the peaks, reach maximum values of about 4.5%. At the equator oscillations are much milder (orange curve): after a first long-duration cycle, the susceptible population at each time settles at a fraction of about 98% of the total while infected people oscillates steadily between 0.06–0.2% of the susceptible population.

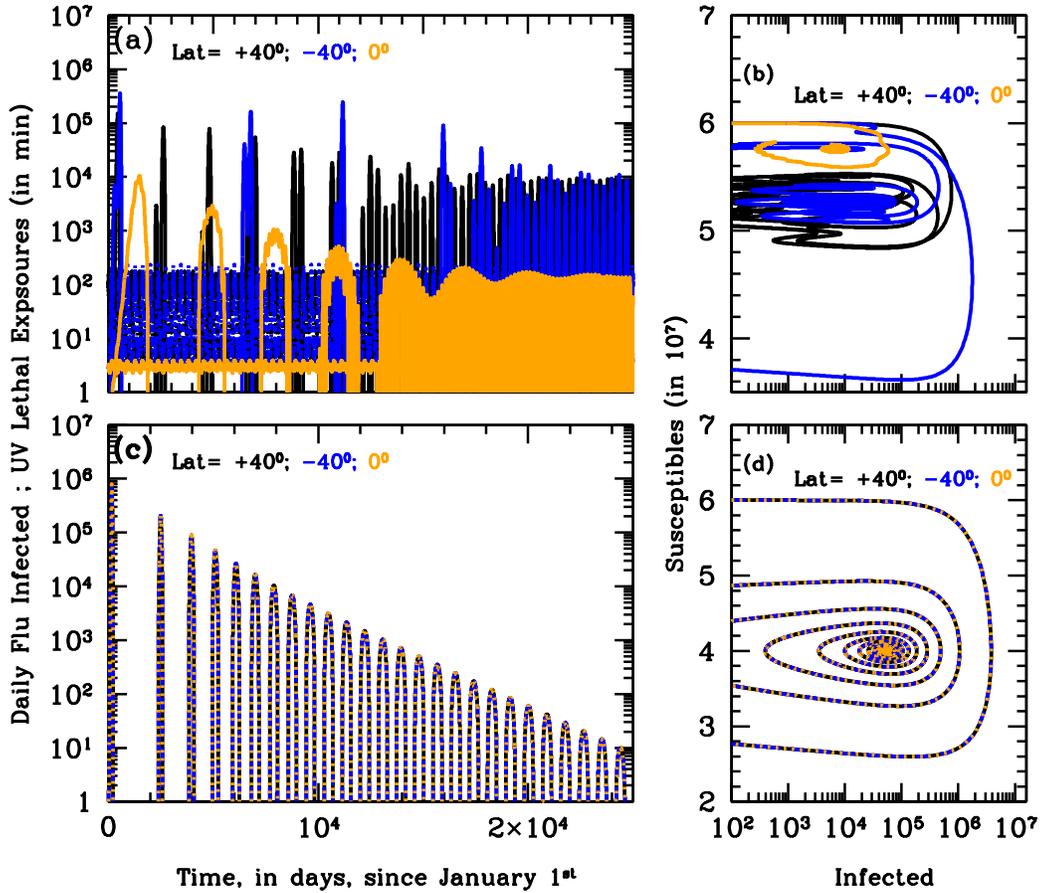

**Figure 4 Slow LOI rate Influenza Epidemics.** Same as Fig. 3, but for a slow LOI rate of $\gamma_{in}$=0.00055. As for the fast LOI rate case, when the solar-pump is active ($\varepsilon$=1; panels (a) and (b)) constructive resonance eventually (in about 30-40 years) builds-in regular sun-modulated oscillations, and the strength of the seasonal outbreaks adjust to quasi-constant values that depend on latitude. However, during the first 30-40 years, intensive outbreaks are sporadic at all latitudes and typically repeat for 2-3 consecutive years in the north and south hemispheres. Without solar-pump ($\varepsilon$=0; panels (c) and (d)) the cycles are the same at all latitudes and are fed by LOIs only till the available number of susceptible is sufficient. The strength of the outbreaks decreases steadily with time, and time-intervals between peaks squeezes, from the inverse of the LOI rate to ~0.2 times that. Susceptible vs actual Infected phase diagrams (panels ((b) and (d)) are plotted for the entire duration of the simulation (100 years). The LOI mechanism alone is able to support a few cycles of the epidemics ($\varepsilon$=0; panel (d)), but these decreases systematically in strength and eventually die out. Switching the solar-pump on ($\varepsilon$=1; panel (b)) allows the system to remain active for the entire duration of the simulation and to stabilize after 30-40 years to susceptibility fractions of ~85% and 97%, at all times, at ±40 degrees and at the Equator, respectively. Correspondingly, the fraction of actual infected over the susceptible population, oscillates steadily between (0.001 – 0.2)% and (0.005 – 0.02)%, at ±40 degrees and at the Equator, respectively.

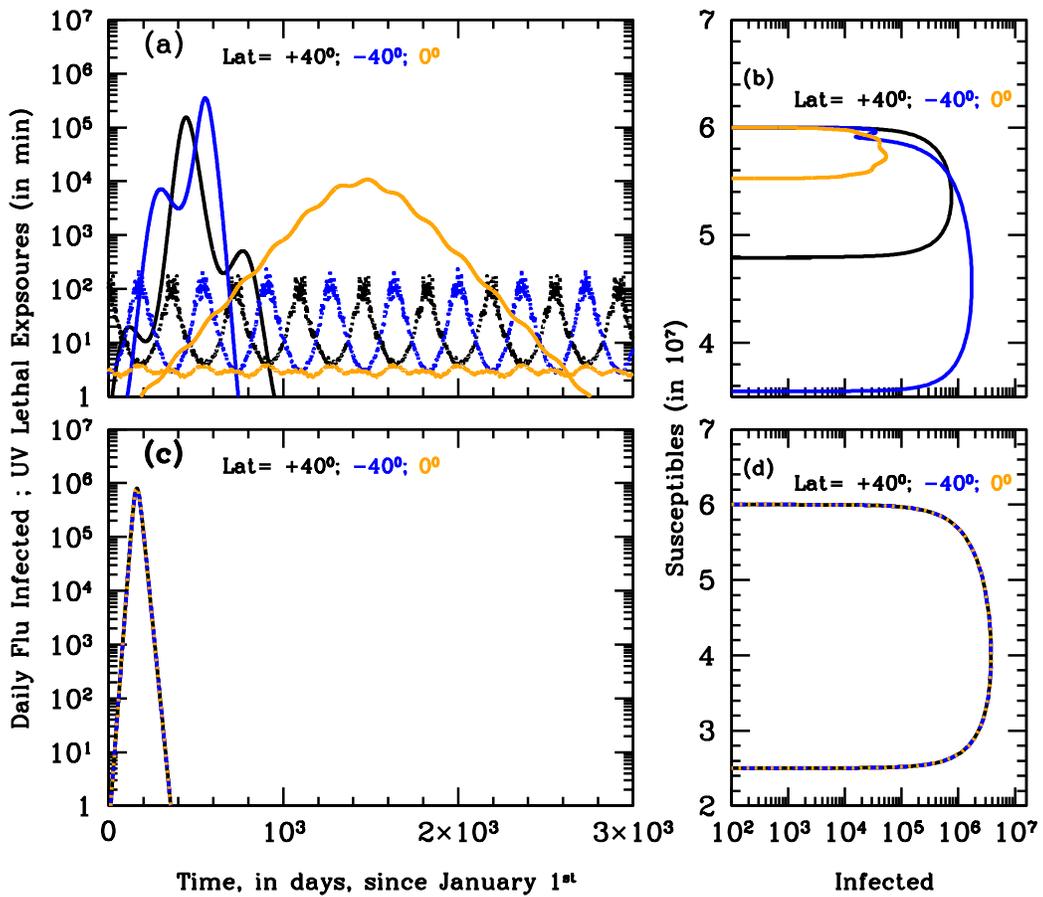

**Figure 5 Non-recurring Influenza.** Same as Figure 3 for the first 8 years (curves of growth of panels (a) and (c)) and 100 years (S–I phase diagrams of panels (b) and (d)) of an influenza-like epidemics with zero LOI rate. When internal dynamics is absent (no LOI) the epidemics cannot become endemic. However, when the solar-pump is active ($\varepsilon=1$; panels (a) and (b)) UV photons from the sun are able to modulate the outbreak and prolong it for a few cycles. A couple of bursts are seen at ±40 degrees, whereas a single long, mild and Sun-modulated epidemics develop at the Equator and last for about 6 years. Without solar-pump ($\varepsilon=0$; panels (c) and (d)) a single intense episode is seen at all latitudes. With epidemics starting during north-hemisphere winters (as in our models), latitude +40 degrees reaches a maximum infected over susceptible fraction of 1%, while this fraction is four times larger at -40 degrees and about ten times smaller at the Equator (panel (b)).

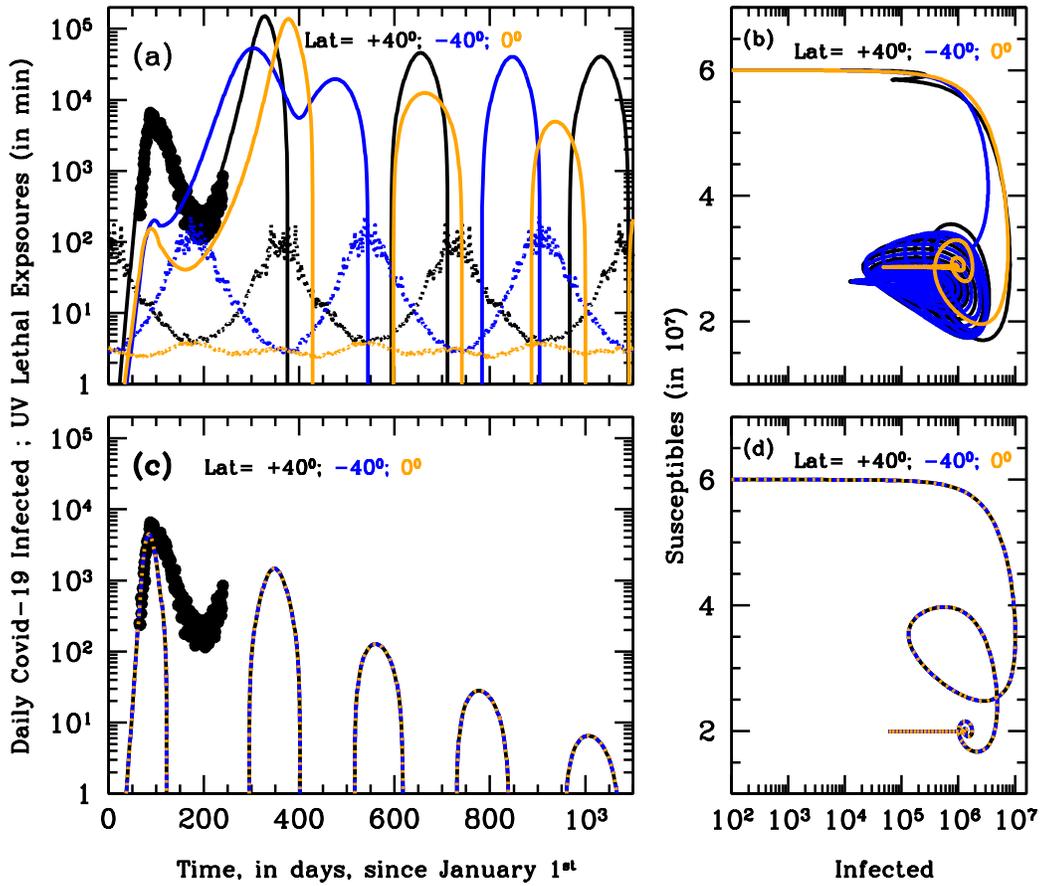

**Figure 6 Evolution of SARS-Cov-2 Pandemics**. Panels (a) and (c): simulation of the curves of growths of the daily new infected for the first 3 years of the SARS-CoV-2 pandemics at three different latitudes on Earth, +40 degrees north (black), -40 degrees south (blue) and the Equator (orange), with ($\varepsilon=1$; panels (a) and (b)) and without ($\varepsilon=0$; panels (c) and (d)) active solar-pump. Model parameters are $t_0$=January 6$^{th}$ 2020, $t_{end}=t_0+100$ years, $R_0=3.0$, $\gamma_{out}=0.1$, $\gamma_{in}=0.0055$, $\mu=0.01$ and external lockdown and phase-2 starting on 11 March 2020 (official lockdown date in Italy) and 17 April 2020 (3 days after the first official reopening date in Italy), with halving and doubling times of 28 and 238 days, respectively. Black points in panels (a) and (c) are SARS-CoV-2 data for Italy from 24 February 2020 through 21 August 2020, smoothed over a 4-day moving average. For the active solar-pump case (panel (a)) all curves of growth of new daily infected have been scaled down by a factor of 6 to let the +40$^0$ curve (black) match the Italian data of the current outbreak: the model closely matches observations. In the non-active solar-pump case (panel (c)) model curves needs to be scaled-down by an unrealistically high factor of 300, and only the climbing phase of the outbreak in Italy can be reproduced. Yearly oscillating dotted curves at the bottom of panels (a) and (c) are UV-B/A lethal-exposure estimated for Covid-19. Panels (b) and (d): S–I phase diagrams for the simulations of panels (a) and (c), respectively. The first 10 years of a possible evolution of the epidemics are plotted. When the solar-pump is not active (panel (d)), the curves are the same at all latitudes and the cycles of epidemics die out in 4 cycles because of lack of supply among the susceptible population. Considering the effect of solar-pump makes the epidemics evolve quickly to endemic at all latitude (panel (b)). After the first 3 cycles, and at latitudes of ±40 degrees (black and blue curves), the susceptible population at each time oscillates between 30–55% of the population, and the instant fraction of infected people over susceptible, at the peaks, reach maximum values of <15%. At the equator the amplitude of oscillations is much smaller (orange curve), the susceptible population at each time settles at a fraction of about 50% of the total while infected people oscillates steadily between 0.2–3.5% of the susceptible population

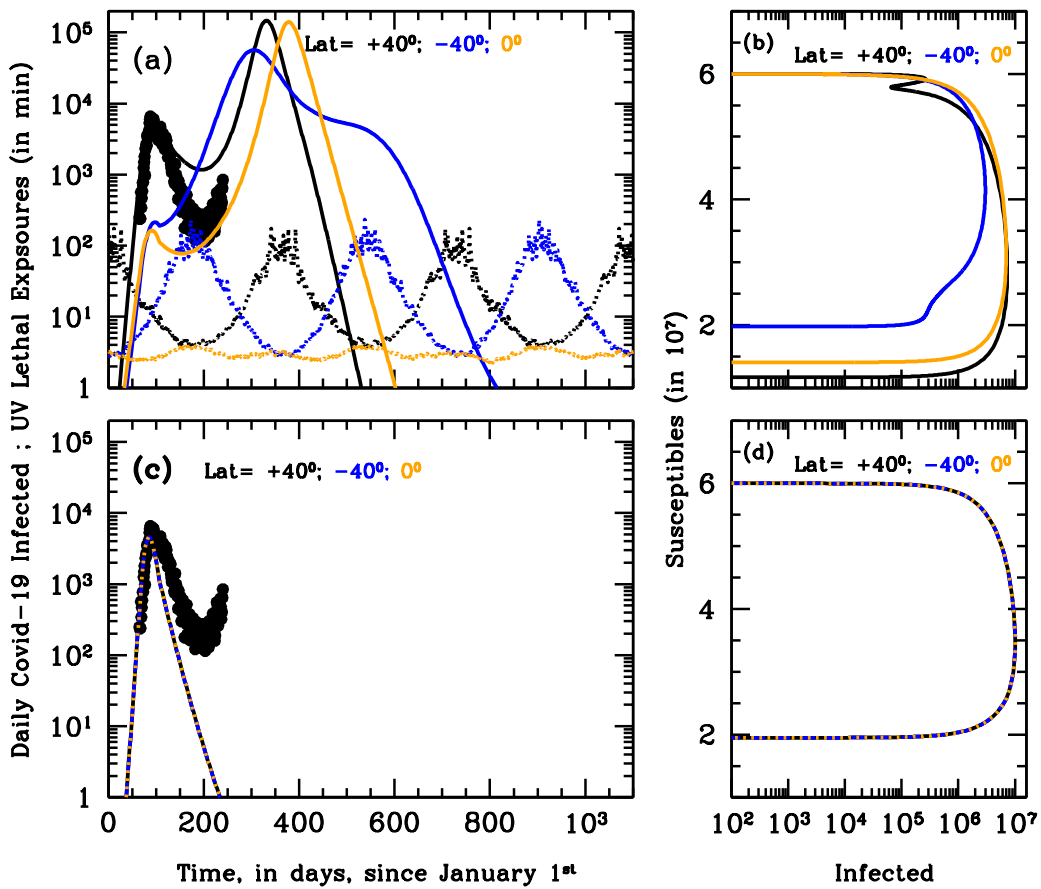

**Figure 7 Secondary wave of no-LOI SARS-CoV-2.** Same as Fig. 6, but for a zero LOI rate ($\gamma_{in}=0$). With these parameters, the model is not able to adequately match the Italian data of the outbreak (black points in panels (a) and (c)), even when the solar-pump is active ($\varepsilon=1$; panel (a)). However, only when UV photons are considered (panel (a)), can the model reproduce the observed differences in geographical spread of the pandemics. Should recovered people preserve their immunity against Covid-19, UV photons from the Sun may still be able to amplify contagions during autumns/winters and produce one additional wave of the SARS-CoV-2 epidemics at all latitudes. With this set of parameters, this cannot happen in absence of UV photons ($\varepsilon=0$; panel (c)), even after lockdown measures are loosened. S–I phase diagrams of panels (b) and (d) are shown for the entire duration of the simulation (100 years). When UV photons are considered (panel (b)), the epidemics last longer (a total of two cycles at all latitudes). With epidemics starting during north-hemisphere winters (as in our models), and in absence of additional lockdown measures, two cycles develop +40 degrees and at the Equator, while a single, longer and asymmetric cycle runs through the end of 2021 in the South hemisphere. The final fractions of susceptible population are 12%, 20% and 35% at +40, 0 and -40 degrees respectively, and maximum 30% instant fractions of infected over susceptible population are recorded in the north hemisphere and at the Equator, towards the end of 2020, while this fraction stays lower (< 11%) at all time during the epidemics in the south hemisphere.

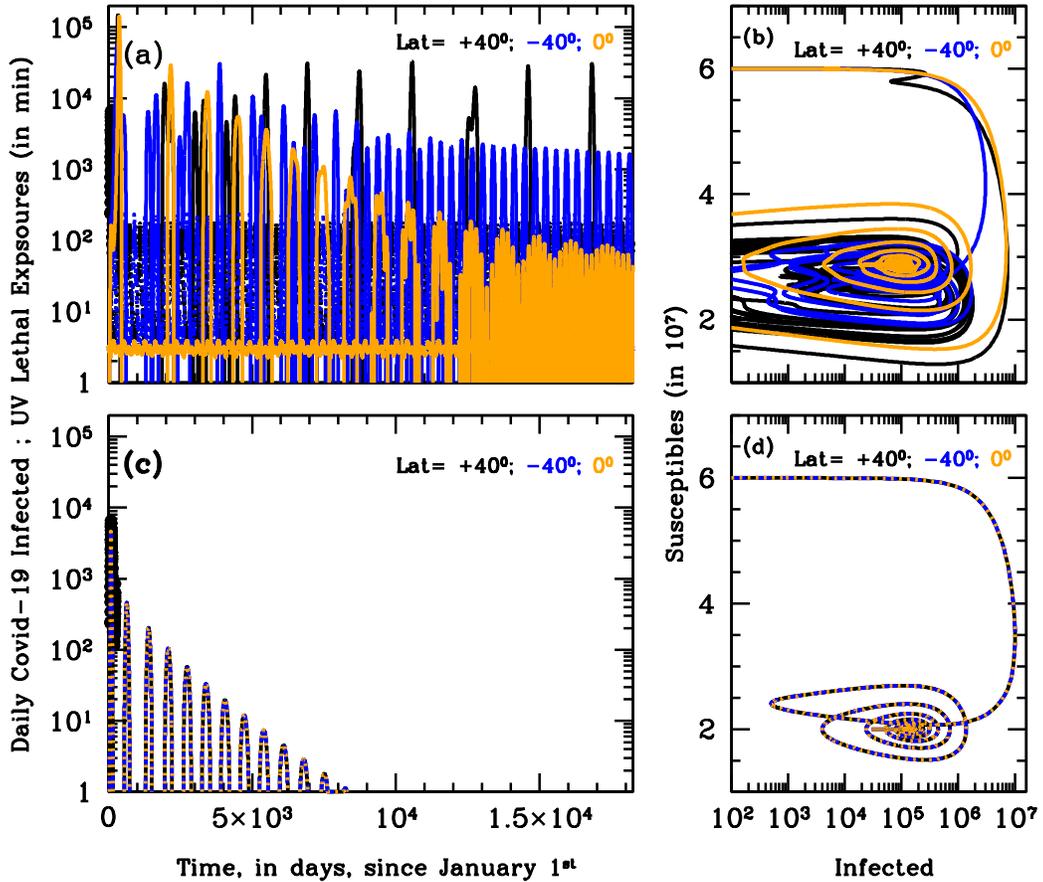

**Figure 8 The nightmare of a recurring long-lasting SARS-CoV-2.** Same as Fig. 6, but for the slow LOI-rate scenario ($\gamma_{in}$=0.00055, or 5-year period) and for the first 50 (panels (a) and (c)) and 100 (panels (b) and (d)) years of the simulation. Both with ($\varepsilon$=1; panels (a) and (b)) and without ($\varepsilon$=0; panels (c) and (d)) solar-pump, cycles proceed driven by the slow-rate reappearing vulnerability of the immune system, with decreasing periods at all latitudes. When UV photons are considered (panel (a)) the initial epidemics' bursts are generally intense and erratic, especially at north and south latitudes, and become steadily recurring and stabilize to the seasonal cycle only after 25-70 years depending on latitude (and in absence of cross-mixing population), fortunately with moderate strength. Should LOI happen on a long timescale (5 years in this simulation), the presence of UV photons from the sun (panels (a) and (b)) would make the first several epidemics bursts in the north and south hemisphere, particularly intense. The fraction of actual people infected at any time over the susceptible population, would reach peak values of up to 30% during the first 25 (south) and 50 (north) years, and would then stabilize, oscillating within the range 0.02–10% (at latitudes ±40) and 0.2–1% (at the Equator), for the next 25-50 years (panel (b)).

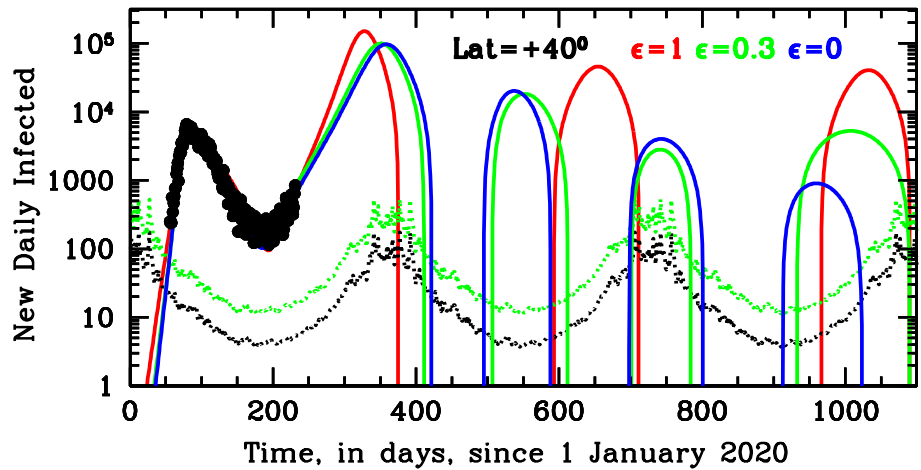

**Figure 9 Evolution of SARS-Cov-2 in Italy for different solar-pump Efficiencies.** Italian daily Covid-19 cases from 24 February 2020 through 21 August 2020 (smoothed over a 4-day moving-average: black points) are shown together with three different SARS-COV-2 model predictions at a latitude of 40 degrees north, for solar-pump efficiencies $\varepsilon=1$ (red curve: same model as in Fig. 6a and 6b), $\varepsilon=0.3$ (green curve) and $\varepsilon=0$ (no solar-pump; blue curve). The three curves differ (other than in solar-pump efficiency) in the values of a number of critical parameters, namely, the epidemic starting time $t_0$ (6, 19 and 22 January 2020 for the red, green and blue curves, respectively), the phase2 starting time (17, 10 and 12 April 2020, for the red, green and blue curves, respectively) lockdown halving times (28, 18 and 15 days, for the red, green and blue curves respectively) and phase-2 doubling times (238, 235 and 210 days, for the red, green and blue curves, respectively). The remaining parameters are common to all simulations, and the same used for Fig. 6–8. All curves of growth of new daily infected have been scaled down by a factor of 6 (as observed) to match the Italian data of the current outbreak. All models closely match observation, however short- and long-term predictions differ significantly in the three cases.

# Supplemental Information

## Transparent Methods

### Our Deterministic Model's Equations

We adopt a simple deterministic SIRS model, in which the Susceptible population (S) interacts not linearly with Infected (I) at a rate β, Recovers (R) or dies (D) with rates $\gamma_{out}$ and μ, respectively, and can become again susceptible (loses immunity) with a rate $\gamma_{in}$:

$$\frac{dS}{dt} = -\beta \frac{S\,I}{N} + \gamma_{in}\,R$$
$$\frac{dI}{dt} = \beta \frac{S\,I}{N} - (\gamma_{out} + \mu)I$$
$$\frac{dR}{dt} = \gamma_{out}\,I - \gamma_{in}\,R$$
$$\frac{dD}{dt} = \mu\,I$$

We wrote a simple Fortran-90 routine that solves this system of differential equations (non-linear only in I and S) iteratively, for arbitrary values of the model parameters. We adopt a time-unit of 1 day and, at each iteration, perform infra-day integrations.

We further introduce three optional external forcing terms that act on the reproductive number $R_t$ (effectively on the contact rate β=β(t)) by attenuating or amplifying it on timescales of 1 day (lockdown and phase-2 terms) and minutes (solar-pump) and may be switched on and off during runs:

$$\beta = \beta_0 (1 - \varepsilon \sin(\alpha)) \begin{cases} 1, & no\ lockdwon/phase2 \\ e^{-(t-t_{LD})/T_{LD}}, & t_{LD} \leq t \leq t_{Ph2} \\ A e^{(t-t_{Ph2})/T_{Ph2}}, & t \geq t_{Ph2} \end{cases}$$

where, ($t_{LD}$, $T_{LD}$), and ($t_{Ph2}$, $T_{ph2}$) are initial and e-folding times of lockdown and Phase-2, respectively (when present), and $A = e^{-(t_{Ph2}-t_{LD})/T_{LD}}$.

### Solar UV Data

In our study, to compute lethal times $\tau_0$, we use Solar UV data made available by the Tropospheric Emission Monitoring Internet Service (TEMIS) archive (http://www.temis.nl). TEMIS is part of the Data User Programme of the European Space Agency (ESA), a web-based service that stores, since 2002, calibrated atmospheric data from Earth-observation satellite. TEMIS data products include measurements of ozone and other constituents (e.g. $NO_2$, $CH_4$, $CO_2$, etc.), cloud coverage, and estimates of the UV solar flux at the Earth's surface. The latter are obtained with state-of-the-art models exploiting the above satellite data.

### SARS-CoV-2 Pandemics Data

We collect SARS-CoV-2 data pandemics for Italy from the on-line GitHub repository provided by the Coronavirus Resource Center of the John Hopkins University (CRC-JHUL https://github.com/CSSEGISandData/COVID-19). Such data are provided daily by the Italian civil-defense department. CRC-JHU global data are updated daily and cover, currently, the course of epidemics in 261 different world countries, by providing the daily cumulated numbers of Confirmed SARS-CoV-2 cases, Deaths and Healings. We use CRC-JHU data from 24 February 2020 (recorded start of the epidemics in Italy) through 21 August 2020.

Black points of Fig. 6–9 are the 4-day moving average of the daily number of confirmed SARS-CoV-2 cases in Italy.